# Morphology and depletion force-based large-scale self-assembly of nanocubes on surface

Yeonhee Lee‡,[1] Seungsang Cha‡,[1] Yuna Kwak‡,[1] Nicholas Juntunen‡,[2] Grant M. Rotskoff,[2]* and Jwa-Min Nam[1]*

[1]Department of Chemistry, Seoul National University, Seoul 08826, South Korea

[2]Department of Chemistry, Stanford University, Stanford, California 94305, USA

‡These authors contributed equally to this work.

*Corresponding authors

## Abstract

Self-assembling nanoparticles is a highly efficient and facile way to form functional nano-, micro- and macrostructures. However, currently available methods lack precision and controllability in size, shape and composition, suffer from poor reproducibility and scalability, and require complex steps and expensive materials. Here, we present the uniform morphology-induced and depletion force-directed nanoparticle assembly on surface (MIDAS) method with gold nanocubes (AuNCs). Using this approach, morphology-sensitive depletion forces trigger shape-selective flocculation and assembly of the AuNCs with uniform size and shape. Importantly, the surface roughness-controlled substrate drives the large-scale formation of AuNC-assembled monolayers (2D AuNAMs) or three-dimensional AuNC-assembled multilayers (3D AuNAMs) in a highly specific manner without any complex ligand modification or preparation steps. Lattice-gas modeling and kinetic Monte Carlo simulations show the depletion force is the key parameter determining supercrystal morphology and corroborate the experimental findings. This work establishes a generalizable nanoparticle assembly mechanism and demonstrates the broad applicability of the MIDAS strategy for scalable fabrication and patterning of 2D and 3D nanoparticle architectures.

## Main

Nanoparticle self-assembly is widely used to fabricate well-aligned nanoparticle superstructures, with unique and collective properties and functions. Precise structural regulation of these structures, including the shape, size, composition, and assembly position, can tune their physical and chemical properties, opening a wide range of possibilities for their applications in optics, sensing, and catalysis.[1-3] Unlike self-assembly systems found in nature, however, the facile, large-scale preparation of nanoparticle assemblies with high structural and compositional controllability remains a fundamental challenge in realizing the potential of assembly-based nanomaterials.

Diverse assembly methods based on modulating the interactions between nanoparticles have been reported, encompassing Watson-Crick base-pairing interactions,[4,5] chemical bonds,[6] van der Waals forces,[7,8] electrostatic forces,[9,10] and depletion forces.[11,12] Each method offers unique advantages and limitations. For example, in the case of chemical ligand-based assembly, high structural and functional controllability can be achieved with the aid of programmable or specially-designed chemical ligands.[13,14] However, as the prerequisite ligand modification chemistry is not universally applicable to all materials, the material compatibility of building blocks is often limited. Furthermore, ligand exchange often introduces time-consuming and laborious steps into the preparation process.

An alternative approach is based on morphology-directed assembly principles, utilizing shape-dependent forces such as van der Waals[7,8] and depletion[11,15] forces. Therefore, topological characteristics can direct the assembly,[16-19] providing relatively simple protocols with high compatibility of building block shapes and materials.[11,20,21] Commonly adopted procedures are based on evaporation or an air-liquid interface,[22,23] but the difficulty of controlling fluid interfaces can lead to low controllability and uniformity of the assembled structures. Previous works have shown that depletion forces were capable of inducing selective flocculation and self-assembly of nanoparticles. However, there have been several limitations and challenges in understanding the mechanism and ensuring controllability, patternability, scalability and general applicability for various systems and conditions.[24]

Here, we report the uniform morphology-induced and depletion force-directed nanoparticle assembly on surface (MIDAS) method that enables the facile and selective formation of 2D nanocube-assembled monolayers (2D NAMs) and 3D nanocube-assembled multilayers (3D NAMs), through the tuning of nanoparticle-substrate interactions via selection of substrate material and surface morphology (Fig. 1). Uniformly synthesized single-crystalline Au nanocubes (AuNCs), surfactant micelles which act as a depletion force-creating depletant, and a solid substrate are basic components in a typical MIDAS system. Using the same AuNCs, 2D AuNAMs are fabricated on highly smooth Si wafer substrates and 3D AuNAMs on slightly less smooth borosilicate glass substrates. The 2D and 3D assembly process at the liquid-solid interface is investigated in real time with optical microscopy, providing further insight into the effect of nanoparticle-substrate interactions on assembly kinetics. Moreover, we harness the sensitivity of depletion interactions to the morphological characteristics of the opposing surfaces[15,25] to directly control the assembly mode. The roughness of the substrate is correlated to the type of final structure through the overlapped volume, which dictates the depletion potential. Combining these experimental results with kinetic Monte Carlo simulations, we develop a framework for understanding the self-assembly of nanoparticles on surfaces.

The MIDAS protocol is a straightforward method for constructing large-scale nanoparticle assemblies among identically sized and shaped nanoparticles, without the use of functional surface ligands or complicated experimental procedures. These AuNC-assembled structures feature high crystallinity and 2–4 nm interparticle gaps that can generate strong and uniform plasmonic coupling over a wide region, enabling their potential use as surface-enhanced Raman scattering (SERS) substrates. Furthermore, MIDAS can open new possibilities for fabricating not only layered assemblies and supercrystals but also pattern-directed self-assembled structures.

## Morphology-induced and depletion force-directed 2D and 3D nanocube assemblies on surface

The AuNC building blocks used for assembly were synthesized by adopting a previously reported method.[26] The seed-mediated, bromide ion-assisted protocol was used to synthesize uniform 50-nm nanocubes having a size variation of 1.3 %. For assembly, a suspension of AuNCs is mixed with benzyldimethylhexadecylammonium chloride (BDAC) solution and deionized water, such that the final concentrations were 1.6 nM AuNC and 55 mM BDAC (Fig. 1). The mixed solution was then dropcast onto the substrate and incubated at 25 °C for 2–3 hours, during which the AuNCs were assembled at the liquid-solid interface. Subsequently, the substrates were washed with a highly concentrated BDAC solution to remove residual unassembled nanoparticles and left to dry under ambient conditions (Fig. 1). BDAC forms micelles acting as depletants to provide depletion potential for nanoparticle assembly, while the highly concentrated washing solution prevents nanoparticle disassembly and redispersion of the assembled superstructures by maintaining sufficient depletion potential. A detailed experimental description can be found in the Methods and Supplementary Notes 1-3.

As described above, the primary assembly force is the depletion potential, which is expressed as,

$$\mathrm{U_{dep}} = -\Delta\Pi \times \mathrm{V_{eff.ov}} \quad (1)$$

where $\Delta\Pi$ is the osmotic pressure and $V_{eff.ov}$ is the effective overlap of the excluded volume. Near a solid surface, there is a volume physically inaccessible to the depletant due to its size, called the excluded volume. The overlap of the nanoparticle-nanoparticle or nanoparticle-substrate excluded volumes is entropically favored,[27] which leads to an attractive depletion force and triggers nanoparticle assembly on the substrate surface (Supplementary Note 4). Because the osmotic pressure rises as micelle concentration increases, control over the micelle concentration can adjust the depletion potential for nanoparticle self-assembly. Under an optimal concentration of 55 mM BDAC for 50-nm nanocubes, the self-assembly of AuNCs is induced. The dependence of overlapped volume on nanoparticle morphology causes nanocubes, having sufficient depletion potential, to undergo self-assembly selectively.[26,28] Ultraviolet-visible (UV-Vis) extinction spectra and transmission electron microscopy (TEM) images before and after assembly revealed that AuNCs were selectively removed from solution for assembly (Extended Data Fig. 1).

The MIDAS process resulted in the formation of 2D AuNAMs or 3D AuNAMs, depending on the substrates on which AuNCs are assembled. The focus of our research is on the pivotal role of the less-studied nanoparticle-substrate interactions in controlling assembly. The nanoparticle-nanoparticle and nanoparticle-substrate interactions are dominated by the depletion interaction, but other interparticle forces, such as electrostatic repulsion and van der Waals attraction also contribute. After the initial nucleation step, subsequent nanocubes would preferentially assemble adjacent to the nucleus, so as to increase coordination with other nanocubes and maximize the total interaction. By the same principle, nucleation of the second layer, in which a nanocube is deposited on top of other assembled nanocubes, is comparatively less favored, as there are fewer neighbors to coordinate to. Relative to the nanoparticle-nanoparticle interaction, the nanoparticle-substrate interaction can be tuned to modulate vertical and lateral growth rates and thus the growth mode (Supplementary Note 5). When the nanoparticle-substrate attractive interactions are comparable to or stronger than the nanoparticle-nanoparticle interactions, lateral growth dominates to form 2D AuNAMs; weaker nanoparticle-substrate attractive interactions instead favor island growth to the form of 3D AuNAMs. This phenomenon is akin to atomic growth at heterointerfaces, where the growth mode is determined by the interfacial energy between the two materials.[29]

The MIDAS method was first tested on Si wafer and borosilicate glass substrates. The formation of 2D AuNAMs was observed on highly smooth Si wafers (Fig. 2a) while 3D AuNAMs were formed on slightly less smooth borosilicate glass substrates (Fig. 2b). Since the difference in depletion force among these two substrates is not significant (the arithmetic roughness parameter ($S_a$) values of Si wafer and borosilicate glass substrates are 0.07 nm and 0.16 nm, respectively), other factors, including electrostatic and van der Waals forces between nanoparticles and the substrates, also contribute to the formation of the assembled structures. The adsorption of BDAC ligands on both the nanoparticles and substrate imparts positive charges that repel one another. The zeta potential of BDAC-capped AuNCs was +55.4 mV, indicative of BDAC adsorption. For a CTAB-capped system, also a quaternary ammonium halide surfactant, it was reported that the zeta potentials for Si and $SiO_2$ were +4 mV[30] and +50 mV[31], respectively, suggesting that there is stronger electrostatic repulsion for glass substrates than Si wafers. Furthermore, the van der Waals force between a nanoparticle and Si wafer should be stronger than the van der Waals force between a nanoparticle and glass substrate, as the Hamaker constant[32] between Au-water-Si (102 zJ) is much larger than that of Au-water-$SiO_2$ (28 zJ). The stronger electrostatic repulsion and weaker van der Waals attraction reduce the net attractive interaction between nanoparticle and substrate on borosilicate glass.

The effect of AuNC and BDAC concentrations on nucleation and growth dynamics was also investigated for both 2D AuNAMs and 3D AuNAMs (Extended Data Fig. 2 & Supplementary Note 6). We found that the nucleation density increases with nanocube concentration while the depletion potential, mediated by the BDAC concentration and the substrate, had a complex effect on nucleation. Multiple regimes were apparent, particularly for 3D AuNAM formation. At the low BDAC concentrations below a threshold, nucleation was not observed as the depletion potential is insufficient to overcome the energy barrier. At the concentrations above this threshold, nucleation has a stronger dependence on BDAC concentration due to an increase in the driving force. The weaker nanoparticle-substrate interaction on a borosilicate glass was also reflected in the reduced nucleation density compared to Si wafers.

The growth of 3D AuNAMs exhibited strong site-selectivity in nanocube deposition. The nanocubes tended to preferentially deposit on kink and step sites so as to maximally coordinate to their neighbors. Thus, growth along the <110> direction to expose {100} facets was favored to form a cubic shape (Supplementary Note 5).

This understanding of the depletion-induced mechanism provides design principles by which the MIDAS method can be broadly applied to different nanostructures with varying sizes and shapes. The MIDAS can be studied with various substrates, depletants and nanoparticles (Fig. 2d-f, Extended Data Fig. 3 and 4, Supplementary Note 3). In particular, the MIDAS method was successfully applied to other flat substrates including template-stripped (TS) Au film ($S_a = 0.3$ nm) and thermal oxide ($SiO_2$) film on Si wafer ($S_a = 0.08$ nm) (Extended Data Fig. 3). That the TS Au films induced 2D AuNAMs and the $SiO_2$ films 3D AuNAMs could be predicted from calculation of nanoparticle-substrate interactions, suggesting the general applicability of this approach.

Structural and optical characterization was carried out for 2D AuNAMs on Si wafers and 3D AuNAMs on borosilicate glass substrates. The average width of the 2D superstructure domains was 11 ± 3 μm, and the surface coverage of the monolayer on the substrate was ~60 % after drying. Scanning electron microscopy (SEM) images revealed that the AuNCs were arranged in a square, close-packed lattice. The lateral interparticle spacing of the 2D AuNAMs was determined to be ~2 nm (Fig. 2a) by subtracting the nanocube size (51.9 nm) from the lattice constant (53.9 nm), measured using the grazing-incidence small-angle X-ray scattering (GI-SAXS) (Fig. 2c, top). The 3D AuNAMs on glass showed cubic crystals (Fig. 2b). The average edge length of the 3D superstructures was 14 ± 3 μm and the average number of the stacked AuNC monolayers was 4 ± 1, with 3–5 layers being predominant (Extended Data Fig. 5b, d). It should be noted that it is possible to create 3D NAM structures with different domain sizes by controlling the BDAC and AuNC concentrations (Supplementary Note 5). The lateral interparticle gap size for the 3D AuNAMs was determined to be 4 nm based on the lattice constant of 55.9 nm obtained from the GI-SAXS (Fig. 2c, bottom). We attribute the difference in lateral interparticle spacing between 2D monolayers and 3D superstructures to steric repulsion, as low-coordination particles have more spatial freedom for ligand deformation, allowing for narrower gaps between particles.[33] The vertical interparticle gap size, obtained with the atomic force microscope (AFM) height measurements and known nanoparticle size, marginally decreased as the number of layers increased (Extended Data Fig. 5b). Electromagnetic simulation results with infinite square lattice nanocube structures matched the experimental reflection spectra, supporting the high crystallinity and the large domain size of the assembled structures. The ultraviolet-visible-near infrared (UV-Vis-NIR) spectrum from monolayer 2D AuNAMs exhibited a resonance peak at 770 nm, induced by a quadrupole gap mode (Fig. 2g, top). For the 3D AuNAMs, plasmonic coupling in the vertical direction generated a more complex set of resonance peaks, as expected from the weighted simulation results (Fig. 2g, bottom and Extended Data Fig. 5c).

***In situ* observation of the growth processes of the 2D AuNAMs and 3D AuNAMs.**

The formation of the 2D and 3D assembled superstructures was further investigated using the *in situ* inverted reflected bright-field light microscopy, which allowed us to directly observe the nucleation and growth processes of the superstructures formed on surfaces in real-time. Due to interference from scattered light by densely dispersed nanocubes in solution, the *in situ* real-time imaging was conducted using transparent substrates (TS Au film and borosilicate glass for 2D AuNAMs and 3D AuNAMs, respectively). Image analysis of assembly dynamics provided further insight into substrate-dependent growth rates and patterns (see Supplementary Note 7 for more details on image processing).

On TS Au film substrates, nanocube nuclei formed immediately following dropcasting, with a nucleation density of 2 per 100 $\mu m^2$ observed within the first minute (Supplementary Video 1, Fig. 3a-(i)). The rapid onset of the nucleation stage suggests a low nucleation barrier and large driving force. Subsequent AuNC deposition preferentially occurred at the edges of stable 2D nuclei rather than generating new nuclei. This separation of nucleation and growth phases arises from both the reduction of bulk solution nanoparticle concentration leading to decreased nanoparticle flux to the surface, and the local consumption of nanoparticles by pre-existing nearby domains.[34] The optical images revealed that during the growth of the nuclei, small nanoparticle domains fluctuated until well-oriented AuNC assembly led to more energetically stable positions (Fig. 3a-(ii)). The continuous process of AuNC toggling and diffusion[35] not only allowed byproducts to be filtered, but also improved the crystallinity of the assembled structures and increased their likelihood of reaching a thermodynamically favorable arrangement. The growth of the assembled structures, as ascertained from their surface coverage, increased linearly (Fig. 3a), reaching monolayer saturation after 40 minutes (Fig. 3a-(iii)).

On the borosilicate glass substrates, the assembly proceeds along a pathway distinct from the formation of 2D AuNAMs (Supplementary Video 2). Initially, a metastable state of small nuclei was observed (Fig. 3b). After an hour, the stabilized nuclei began to rapidly grow to form a cuboid-shaped first layer (Fig. 3b-(i) and (ii)). In contrast to the growth of 2D AuNAMs, the relatively weaker interactions between AuNCs and the substrate increased the nucleation barrier, significantly reducing the nucleation rate and the number of nucleation sites. At 80 minutes and 106 minutes, secondary and tertiary nucleation was visible in greenish-yellow and orange, respectively (Fig. 3b-(iii) and (iv)).

The direct visualization of nanoparticle growth points to guidelines for superstructure construction. The net interaction inducing vertical nucleation was weaker than that for lateral growth due to the additive nature of nanoparticle interactions. Thus, only three such nucleation events occurred over the course of the growth stage, though on the surface, thousands of nanocubes attached laterally. The same logic underpinned the observation of layer-by-layer-like growth on the nanocube layer – once a nucleation site is formed, there is rapid propagation of the additional layer. In fact, the growth of the second layer was faster than that of the first, as AuNCs would show stronger attractive interactions with preformed nanocube layers (nanoparticle-nanoparticle) than the glass substrate (nanoparticle-substrate).

We then monitored the drying of assembled structures after assembly. Drying occurred under ambient conditions after washing with a highly concentrated BDAC solution to remove excess unassembled nanoparticles. Removal of water from the interparticle distance and contraction of the ligand shell[36] causes the surface coverage of 2D AuNAM to be reduced to 73.4% (Fig. 3c) and that of 3D AuNAM to 73.6% (Fig. 3d) of their solution-state values. From the dried gap distances and the degree of shrinkage, we calculated the interparticle distances prior to drying to be 10.8 nm and 13.0 nm, respectively.

**Substrate surface roughness-controlled AuNC assemblies**

Next, a substrate surface roughness control-based strategy was implemented to direct specific superstructure growth modes. This strategy was based on the sensitivity of the depletion potential to the effective overlapped excluded volume (equation (1)), which in turn should be dependent on the geometries of the interfacing structures. By controlling the overlapped excluded volume between AuNCs and the substrate via the surface roughness of the solid substrate, 2D NAMs or 3D NAMs were exclusively formed (Fig. 4a). A series of Si wafers with arithmetic roughness parameter ($S_a$) values ranging from 0.07 nm to 13.4 nm were obtained by plasma etching with $SF_6$ gas under different radio frequency power values (Supplementary Note 8). The phase diagram of assembly under different values of surface roughness and BDAC concentrations is presented in Fig. 4b. The roughness value was correlated to $U_{NP-sub}$, while the concentration of BDAC affected both $U_{NP-sub}$ and $U_{NP-NP}$.

As the surface roughness increased, a higher BDAC concentration was required to induce assembly formation. For $U_{NP-NP}$ = 6$k_B$T, a transition was observed from 2D monolayers to 3D superstructures upon increasing surface roughness, due to a reduced $U_{NP-sub}$. Monolayer 2D assembly occurred at $S_a$ values less than 2 nm (Fig. 4e), while crystalline cubic 3D assemblies were preferentially formed for $S_a$ values greater than 4 nm (Fig. 4f). Interestingly, tilted superstructures were observed with rough surfaces ($S_a$ = 13.44 nm) and strong depletion potential conditions, due to homonucleation in the bulk solution (Extended Data Fig. 6).

Representative flat- and rough-surfaced wafers, having $S_a$ values of 0.06 nm and 4.48 nm, respectively (Fig. 4g, h), were chosen for more extensive analysis of the influence of surface roughness. Calculation of depletion potential at each point converted their AFM height maps into depletion potential maps for AuNCs interacting with the flat and rough Si wafer at 60 mM BDAC (Supplementary Note 9, Extended Data Fig. 7a). It should be noted that the formulation of overlapped volume takes electrostatic interactions between charged micelles and between charged micelles and charged surface ligands into consideration. It is clear that an ultra-flat surface can maximize the overlapped volumes (Fig. 4k). The flat Si substrate showed no significant peaks or deep holes, leading to uniformly strong depletion attractions between AuNCs and the substrate ($U_{ave,dep}$ = −5.9 $k_B$T, $U_{std,dep}$ = − 0.05 $k_B$T), as shown in Fig. 4i and Extended Data Fig. 7b, comparable to an ideal flat surface ($U_{dep}$ = −6.0 $k_B$T). In contrast, the rough-surfaced substrate featured various cavities and ridges between AuNCs and the substrate, greatly reducing the overlapped volumes (Fig. 4l). Consequently, this reduced the depletion potential and broadened the overall distribution ($U_{ave,dep} = -1.6\ k_BT$, $U_{std,dep} = -1.1\ k_BT$) (Fig. 4j, Extended Data Fig. 7c). The surface heterogeneity is expected to affect the location of stable nuclei at the nucleation stage, with relatively flat regions favored for nucleation. As BDAC micelles have an average size of 3 nm, characterized by small-angle X-ray scattering (SAXS) and an effective micelle radius of ~5 nm, comparable to the ~4 nm $S_a$ of the rough Si wafers, even nanoscale features can exhibit a substantial effect on the depletion-based interaction. If the size of the depletants increases to be much larger than that of the surface features, a larger portion of the overlapped volume will remain unaffected by these surface features, reducing their overall impact on depletion interactions (Supplementary Note 10).

## Lattice-gas simulations match *in situ* observations

In a real system, van der Waals, electrostatic, and many other forces work together with depletion forces. With kinetic Monte Carlo (KMC) lattice-gas simulations, we were able to separately assess the effect of depletion force with different substrate morphologies on superlattice growth modes (Fig. 5a, Supplementary Note 11). Smoother substrates showed mainly 2D monolayer nanocube growth whereas rough substrates exhibited greater 3D cuboid island growth of nanocubes (Fig. 4d and Fig. 5b). We also captured the separation in assembly timescales. With few attractive substrate sites on rough surfaces, nanoparticles underwent transient deposition and removal until successfully building on established nucleation sites, resulting in 2 orders of magnitude longer assembly time for 3D NAM compared to 2D NAM (Fig. 5d). This qualitatively agrees with the experimental results shown in Fig. 3a, b. As the simulation is agnostic to chemical identity, this supports that surface morphology-driven differences in depletion potential are sufficient to drive the selective growth of 2D or 3D NAMs.

Modeling nanoparticle orientation also revealed grain boundaries between separate nucleation sites in smooth substrate growth observed experimentally. Independent domains nucleated with different orientations experienced an orientation mismatch upon meeting, resulting in grain boundaries (Fig. 5c). In simulations with rough substrate morphology, weaker surface interactions required greater nanoparticle alignment and interaction strength for growth to occur. This also drives assembly self-selection as described above to favor well aligned nanocubes. Layer-by-layer growth consistent with experiment was also observed, though KMC simulations exhibited more vertical layers in 3D assemblies (Fig. 5e, f).

## Mean field theory captures surface-mediated morphology

We developed a mean field theory to further elucidate the observed behavior in both simulations and experiments (Supplementary Note 12). As illustrated in Fig. 4c, this mean field theory qualitatively agrees with our KMC simulations, exhibiting a progressive transition across surface interaction strengths. Notably, the no-assembly region in the KMC simulations match the thermodynamically inaccessible regions in the mean-field theory (Extended Data Fig. 8). Our analysis indicates a preference for deposition on a substrate surface at low surface roughness, while higher roughness favors deposition atop other nanoparticles, consistent with experimental 3D cuboid assemblies. Consequently, we concluded that the nanoparticle-substrate interaction strength is a critical determinant in differentiating between 2D and 3D assembly. Specifically, when $U_{NP-sub}$ is sufficiently high, the system preferentially promotes substrate coating; conversely when $U_{NP-sub}$ is low, $U_{NP-NP}$ will dominate and drive deposition atop existing nanoparticles.

## Patterned MIDAS through surface morphology control

Using the roughness-dependent phenomena described above, we were able to selectively assemble AuNCs on patterned substrates (Fig. 6a). The patterned substrates were prepared with focused ion beam (FIB) milling on a 50 nm Au film deposited on a Si wafer (Supplementary Note 13). The FIB milling removes Au to expose the Si surface, leaving gold in the desired patterns. The exposed Si surface was then roughened with plasma etching using $SF_6$ and $O_2$ gas. After etching, the residual gold film was removed with potassium iodide solution, exposing the flat-surfaced patterns surrounded by roughened regions. This creates the localized potential wells for nanoparticle-substrate interactions, leading to selective array formation on flat surfaces. AFM height profiles clearly show the roughness difference between the on-pattern and off-pattern regions (Fig. 6c). Two different types of designs were prepared: an array of 2 μm square shapes (Fig. 6b) and the letters 'NAM' with more complex structural features (Fig. 6d). A single-domain 2D AuNAM was formed on each individual island of the array. The size of the patterns affects the growth of assembly (Supplementary Note 14). The selectively assembled AuNC pattern matched well with the shape, size, and position of the initial NAM pattern in both the experimental results and the KMC simulations (Extended Data Fig. 9), demonstrating the potential applicability of MIDAS for nanofabrication by simply controlling morphological and surface roughness factors.

## Scalable MIDAS for uniform SERS substrates

Another advantage of MIDAS is the ease with which it is scalable; various sizes of 2D NAMs were formed by adjusting the droplet volume during dropcasting (Extended Data Fig. 10a, b). For a wafer-scale demonstration, a 4-inch Si wafer was immersed in the AuNC and depletant mixture solution. The bright and uniform gold color of 2D NAMs was observed across the entire wafer surface (Fig. 6e) and SEM images revealed well-assembled 2D NAMs on the surface (Extended Fig. 10e). Both UV-Vis-NIR reflectance and SERS spectra from the 2D NAM wafer were measured to validate structural and optical uniformity. A series of reflectance spectra obtained from ~1 cm diced 2D NAMs on the wafer consistently showed the characteristic plasmonic peak (Extended Data Fig. 10d). Additionally, the SERS mapping performed after Raman dyes were added to the 2D AuNAM structure produced reproducible, uniform SERS spectra (Extended Data Fig. 10c) over a ~2x2 $cm^2$ area with a coefficient of variation of 13.2% (Fig. 6f). These results suggest our MIDAS approach is scalable and could be useful for producing large-area nanoparticle assembly films in a reliable manner.

## Conclusion

In this article, we report the uniform particle morphology and depletion force-based MIDAS method that allows for large-scale 2D or 3D supercrystal growth of nanocubes based on modulation of the nanoparticle-substrate interactions. Morphology plays a vital role, as a selective flocculation facilitates the assembly among nanocubes of uniform size and shape. Furthermore, experimental and simulation results indicate that the substrate roughness can be used as a lever for adjusting nanoparticle-substrate interactions to tune the growth mode between layer and island growth. The assembled crystalline superstructures showed plasmonically coupled nanogaps, expected to be useful for plasmonic nanogap-based applications such as surface-enhanced Raman scattering and strong light-matter interactions.[37] Importantly, it was shown that wafer-scale 2D NAMs can be formed and AuNC monolayers assembled on various patterns. The assembly of AuNCs on patterns were facilitated simply by controlling surface roughness – AuNCs were assembled specifically on smooth surfaced patterns. Together with the high material compatibility with minimal need for surface and ligand chemistry, the MIDAS addresses the lack of simplicity, versatility, spontaneity and scalability in artificial assembly systems and offers a generally applicable and versatile assembly principle that can be widely utilized for preparing assembled materials, platforms and devices useful for plasmonics, photonics, electronics, sensing, imaging and catalysis.

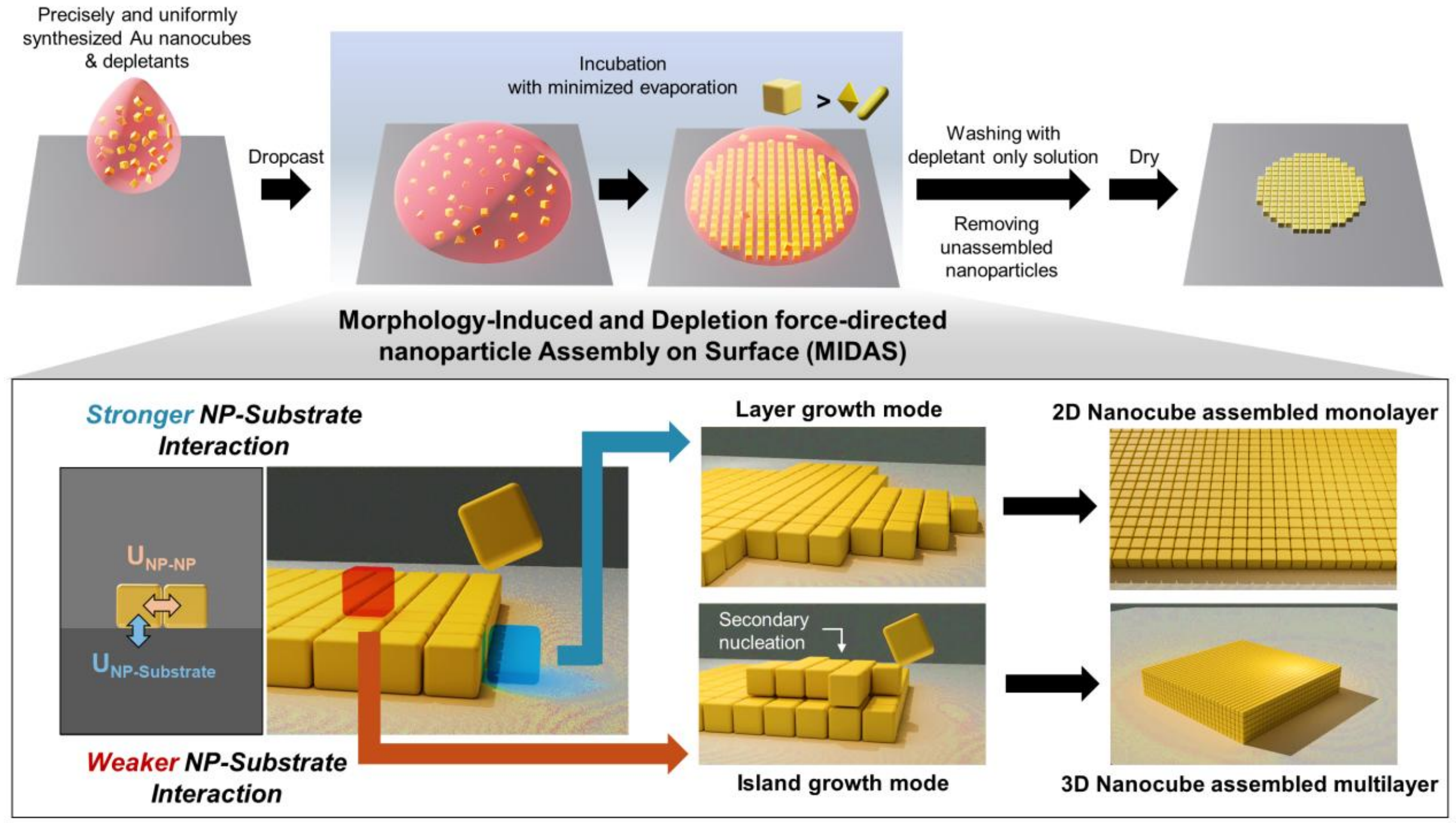

**Fig. 1 | Morphology-induced and depletion force-directed nanoparticle assembly on surface (MIDAS).** The mixture containing Au nanocubes (AuNCs) and depletants is dropcast on the substrate and incubated for a few hours. $U_{NP\text{-}NP}$ denotes the interaction potential between nanoparticles while $U_{NP\text{-}substrate}$ denotes the interaction potential between a nanoparticle and a substrate. The stronger attractive interaction between a nanoparticle and a substrate compared to the interaction between nanoparticles induces the faster lateral growth of AuNCs, leading to the monolayer growth. In contrast, the weaker attractive interaction between a nanoparticle and a substrate leads to the relatively slower growth, inducing the secondary nucleation and layer-by-layer growth on the first AuNC layer, resulting in the island growth mode to create the 3D AuNC supercrystals. After incubation, the remaining supernatant is washed out, and the substrate is dried at room temperature for characterization and further study.

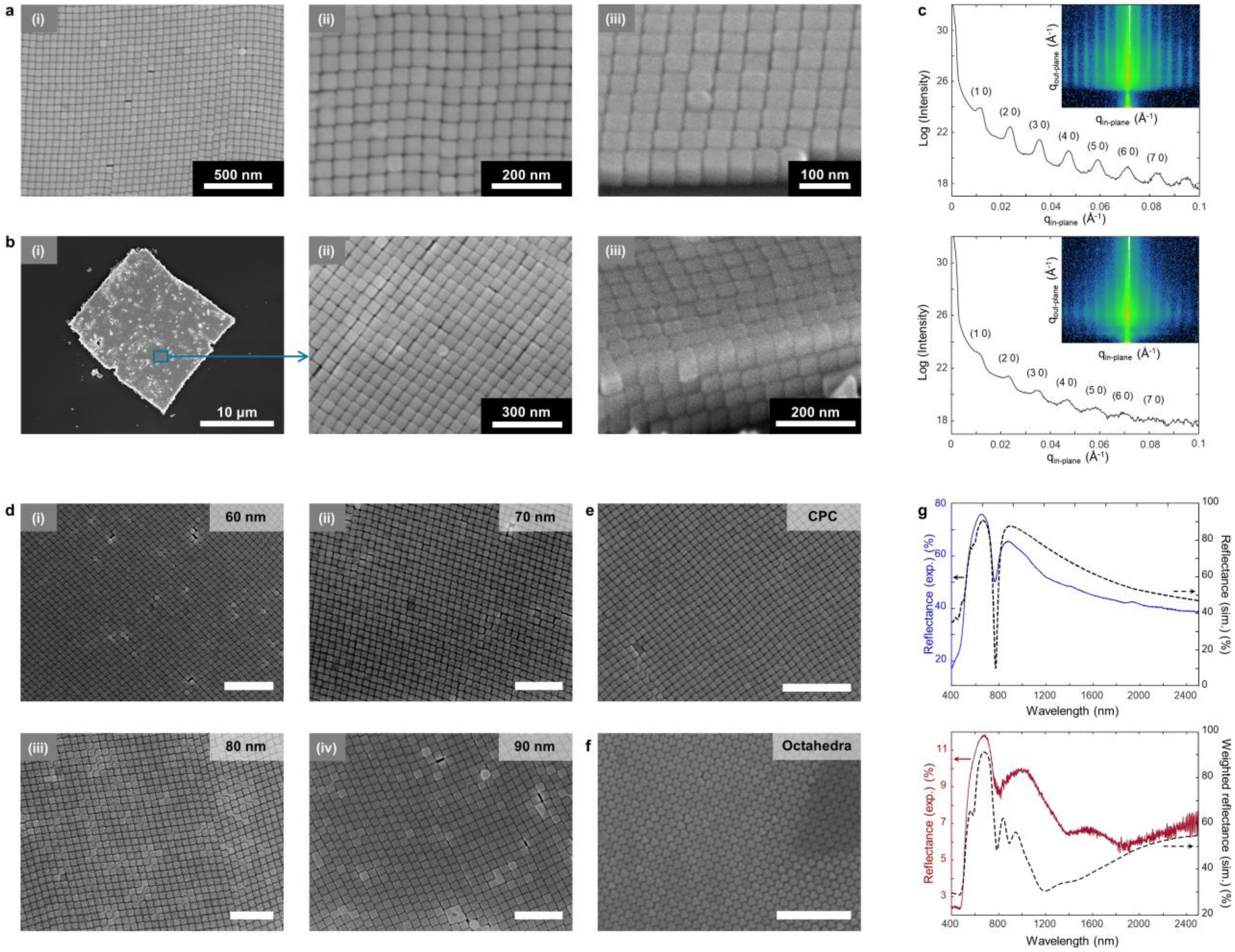

**Fig. 2 | Characterization of the 2D nanocube-assembled monolayers (2D NAMs) and 3D nanocube-assembled multilayers (3D NAMs). a-b**, Scanning electron microscopy (SEM) images of the assembled AuNCs on surfaces. SEM images of the 2D NAMs (**a**) and 3D NAMs (**b**), with tilted images showing 2D monolayer (**a-**iii) and 3D supercrystal (**b-**iii) formation. **c,** The grazing-incidence small-angle X-ray scattering (GI-SAXS) data showing crystallinity of the structures of 2D (top) and 3D (bottom) assemblies, respectively. **d,** 2D NAMs with varying sizes of AuNCs, 60 nm (i), 70 nm (ii), 80 nm (iii), and 90 nm (iv). **e,** 2D NAM formed using cetylpyridinium chloride (CPC) as the depletant. **f,** MIDAS-based formation of Au nanooctahedron-assembled monolayer. Scale bars represent 500 nm (**d-f**). **g,** UV-Vis-NIR reflectance data (solid lines) together with electromagnetic simulation data (dotted lines) of 2D (top) and 3D (bottom) assemblies, respectively.

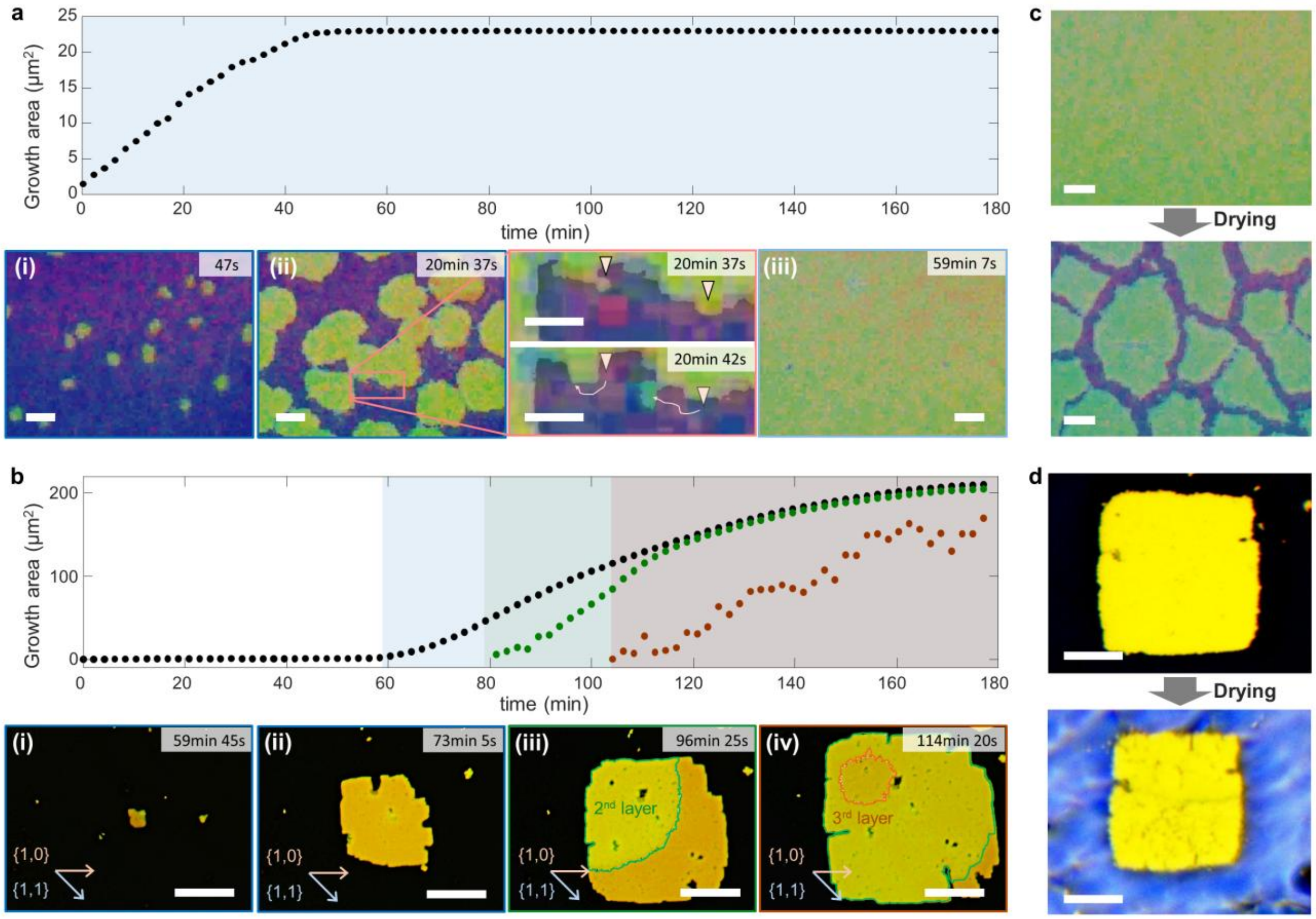

**Fig. 3 | *In situ* monitoring of the MIDAS processes of AuNCs with a bright-field microscope. a,** The average growth area of the two-dimensional assembled structure as a function of time using an *in situ* image sequence. The nucleation and growth stages were captured by microscopic snapshots. Nanoparticles nucleate (**i**) on the substrate within one minute after initiating the assembly. Subsequently, the assembly process undergoes lateral growth via deposition and diffusion (**ii**), resulting in the complete filling of the substrate by the monolayer (**iii**). **b**, Growth pattern of the entire three-dimensional assembled structure (black), the second layer (green), and the third layer (red) over time. Shaded regions represent the metastable stage (gray), growth of the first layer (blue), growth of the second layer (green) and growth of the third layer (red), respectively. Microscopic snapshots show faceted 3D assembly process. Following a metastable state, disordered and unstable nuclei transition into stable nuclei, (**i**) initiating the formation of the first layer with a rectangular structure (**ii**). After the growth of the first layer, the second (**iii**, green solid line) and the third (**iv**, orange solid line) layers are sequentially formed with a time lag. **c-d**, Observation of 2D (**c**) and 3D (**d**) NAM during drying process. Scale bars represent 5 μm.

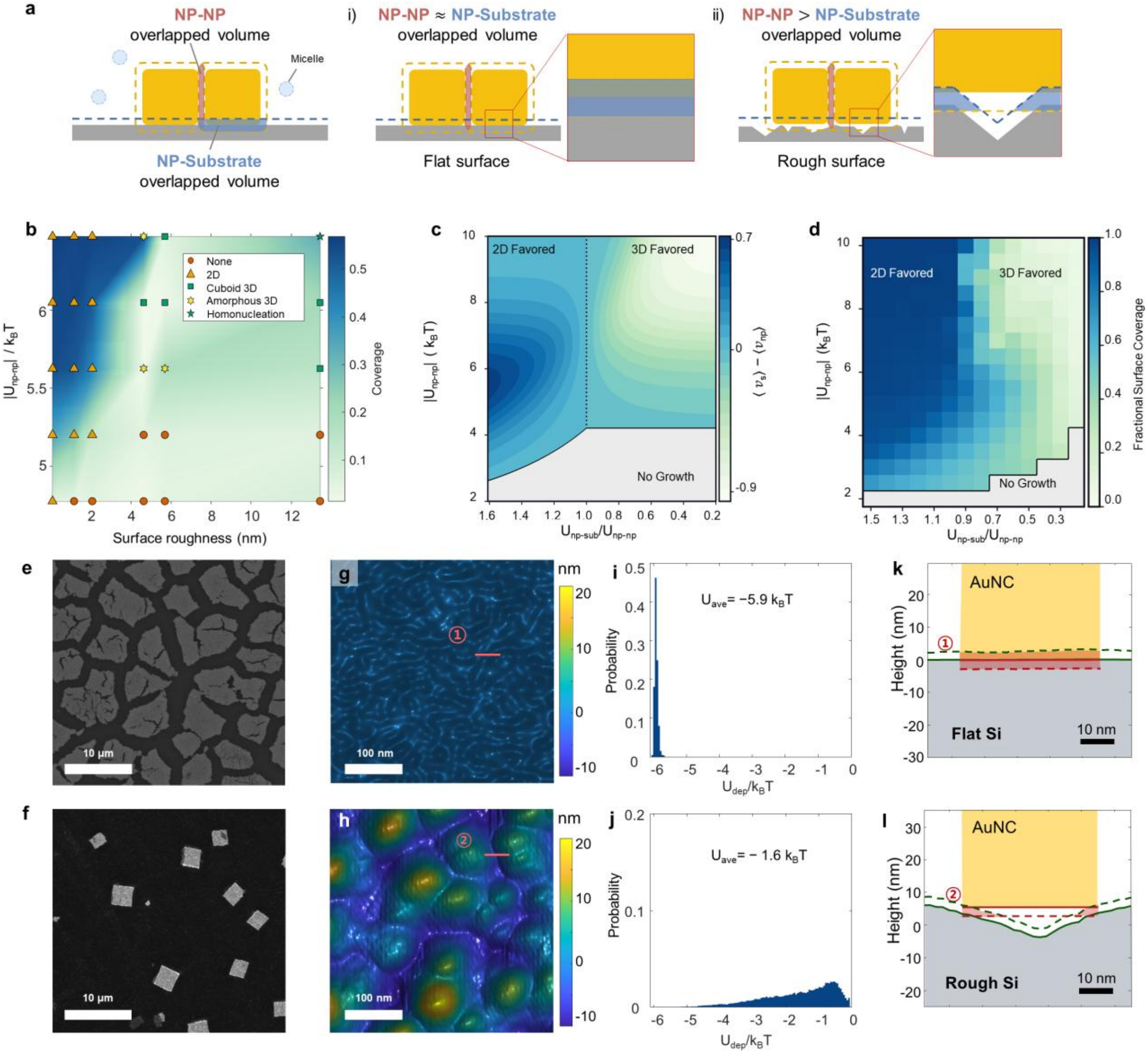

**Fig. 4 | Surface roughness effect on the MIDAS. a,** Schematic figure showing how the overlapped volumes are affected by surface roughness. **b,** A phase diagram of the nanocube assemblies depending on depletion potential strength and surface roughness. **c,** A phase diagram from the mean field theory across a broad parameter range. Kinetic preference is plotted in the blue-to-green scale, and thermodynamic "No Growth" area is shown in gray. **d,** A phase diagram of the assemblies from kinetic Monte Carlo (KMC) simulation parameter sweep. **e-f,** SEM images of the 2D assemblies formed on a flat-surfaced Si wafer ($S_a$ = 0.07) (**e**) and 3D assemblies formed on a rough-surfaced Si wafer ($S_a$ = 4.62) (**f**). **g-h,** Atomic force microscopy (AFM) height profiles of a flat Si wafer (**g**) and a $SF_6$-etched rough Si wafer (**h**). Based on the AFM height profiles, depletion potentials were calculated by the sequential quadratic programming (SQP) algorithm. **i-j,** The depletion potential distribution on the flat Si wafer (**i**) and the rough Si wafer (**j**), respectively. **k-l,** The representative optimal orientations of nanocubes on flat (**k**) and rough (**l**) Si surfaces. Each position is marked in (**g**) and (**h**). The rough Si surface showed a reduced overlapped excluded volume compared to the flat Si surface.

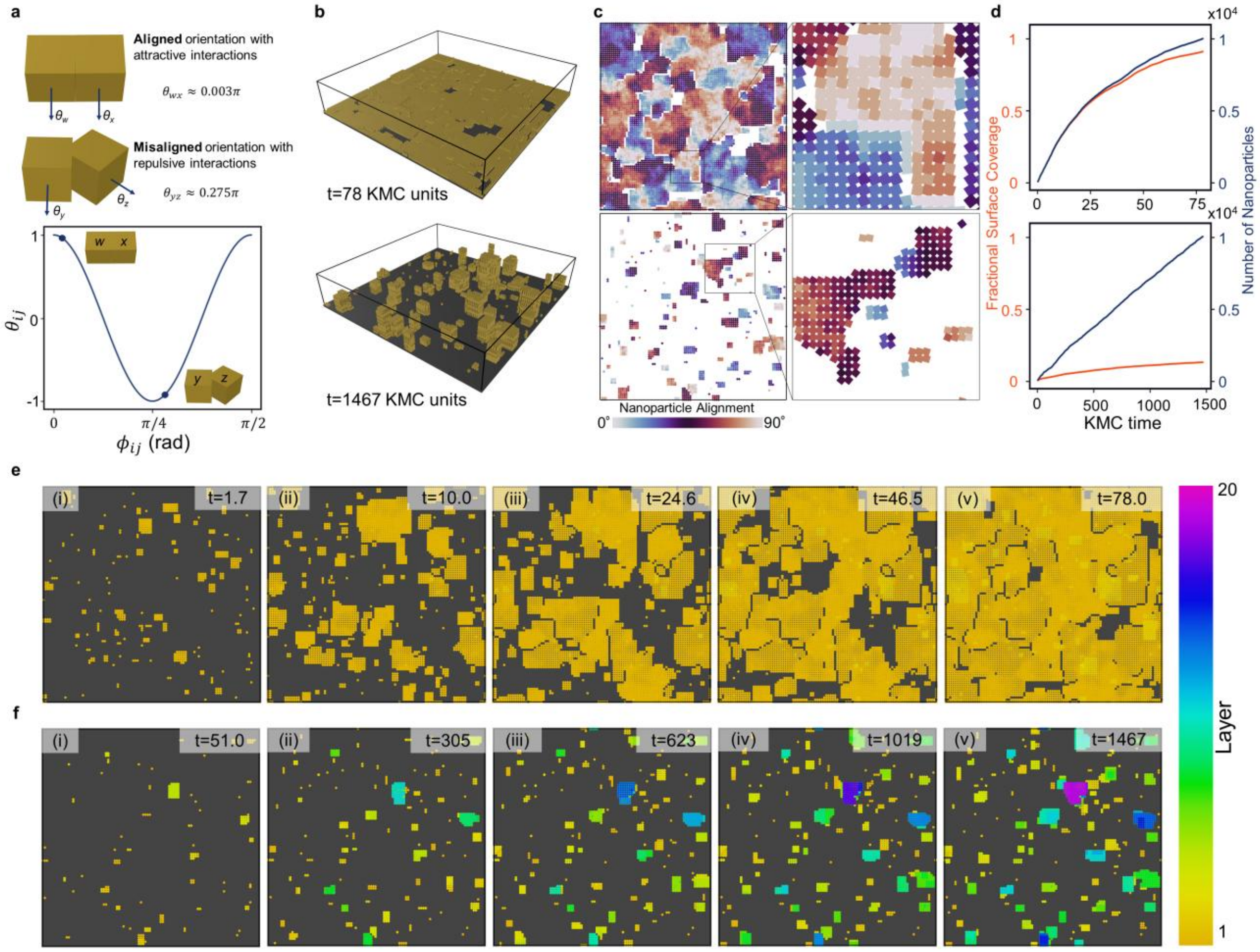

**Fig. 5 | Kinetic Monte Carlo (KMC) simulations of surface morphology-controlled self-assembly of nanocubes. a,** Schematic of the impact of neighboring nanoparticle orientation on interaction strength. **b,** Representative lattice-gas self-assembly modeling for flat (Fig. 4i) and rough (Fig. 4j) surface statistics. **c,** Orientation of nanoparticles deposited on the substrate. Similar colors are aligned, and dissimilar colors are misaligned. **d,** Growth kinetics from simulation. Data was tracked to 10,000 nanoparticles, with the gap between number of nanoparticles and fractional surface coverage indicating the degree of 3D stacking. **e-f,** Simulation trajectories for smooth **(e)** and rough **(f)** surfaces. Times are reported in KMC units. In the smooth system, nanoparticles cover the surface and see minimal vertical stacking ($z_{max}$ = 3). In the rough system, clusters are nucleated and grow vertically to form cuboid structures ($z_{max}$ = 20).

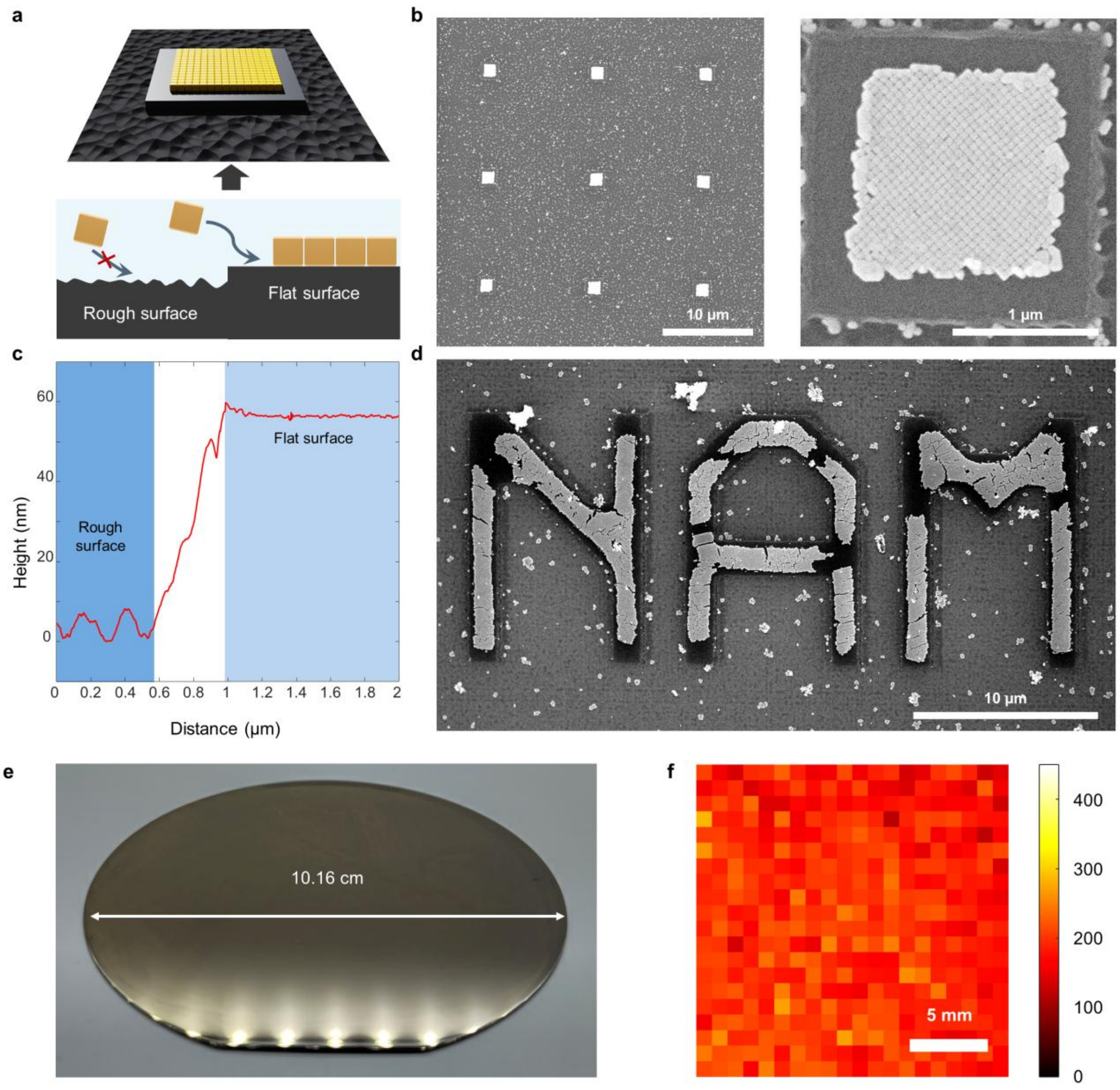

**Fig. 6 | Patterned MIDAS through surface morphology control and scalable MIDAS. a,** Schematic figure of the assembly of 2D AuNAMs on flat-surface patterns. **b,** SEM images of the assembled AuNCs on 2 μm square array patterns**. c**, The AFM line profile of flat and rough surfaces on pattern. The profile includes the edge of the pattern, showing both flat and rough surfaces on and off the pattern, respectively. **d,** SEM image of 2D AuNAMs assembled on the flat-surfaced letter pattern 'NAM'. **e-f,** Large-scale 2D AuNAM assembled on a 4-inch Si wafer (**e**) and a large-area SERS mapping image (**f**) of the AuNAM on the Si wafer, using Raman intensity at 1174 $cm^{-1}$. The samples were incubated in 100 μM crystal violet solution and dried.

## Methods

### Materials

Gold(III) chloride trihydrate ($HAuCl_4 \cdot 3H_2O$, Sigma-Aldrich, >99.9%), silver nitrate ($AgNO_3$, Sigma-Aldrich, >99.9999%), copper(II) nitrate hemipentahydrate ($Cu(NO_3)_2 \cdot 2.5H_2O$, Acros Organics, >98%), hexadecyltrimethylammonium bromide (CTAB, Sigma-Aldrich, >99%), hexadecyltrimethylammonium chloride (CTAC, Tokyo Chemical Industry, >95.0%), benzyldimethylhexadecylammonium chloride (BDAC, Sigma-Aldrich), L-ascorbic acid (Sigma-Aldrich, reagent grade), sodium borohydride ($NaBH_4$, Tokyo Chemical Industry), sodium bromide (NaBr, Samchun Chemicals, >99.0%), cetylpyridinium chloride (CPC, Sigma-Aldrich), gold etchant, standard (Sigma-Aldrich), crystal violet (Sigma Aldrich) silicon wafer (Omniscience, n-type, single-side polished, <100> orientation, prime grade), borosilicate cover glass (Marienfeld, Cat. No. 0101060), silicon oxide wafer (Taewon Scientific, n-type, <100> orientation, prime grade, silicon oxide thickness 1000 Å), gold pellet (Taewon Scientific, 5N), Norland Optical Adhesive 61 (NOA61, Norland Products). All the water used in this work was deionized water (DIW, Milli-Q, >18.0MΩ).

### Synthesis of nanoparticles

AuNCs were synthesized based on a method from the literature,[26] scaled-up 20-fold. AuNCs with 50 nm edge length were synthesized in a 500 mL round-bottom flask. Reactions were conducted at room temperature. First, 368.04 mL of DIW, 120 mL of 200 mM CTAC, 1.2 mL of 20 mM NaBr, 0.7 mL of 10 nm Au nanosphere seeds, and 1.56 mL of 100mM ascorbic acid were mixed with a stirring bar at 500 rpm. Next, 6 mL of 20 mM $HAuCl_4$ solution was added in one-shot and the solution incubated for 2 hours under 500 rpm mixing. The solution was transferred to conical tubes, centrifuged and redispersed twice (3,200 g, 10 minutes) in DIW, and combined into a single conical tube for a final volume of 20 mL to be used as the nanocube stock solution.

Experimental details and protocols for synthesizing Au nanocubes of different sizes, as well as synthesis of Au octahedra and Ag nanocubes can be found in the Supplementary Note 1.

### Substrate preparation

**Substrate cleaning.** Flat silicon wafers, borosilicate cover glasses, and silicon oxide wafers were all prepared following the same procedure: sequential immersion and sonication in acetone, ethanol and DIW for at least 30 minutes in each solvent, followed by drying in inert gas. Cleaned substrates were stored under vacuum conditions until further use.

**Surface roughness-controlled silicon wafer preparation.** Clean Si wafers were treated with plasma etching to control the surface roughness under conditions of 50 sccm of $SF_6$ gas and with the designated radio frequency power for 5 minutes. The arithmetic roughness parameters corresponding to each radio frequency power is described in Supplementary Note 8.

**TS Au film preparation.** Smooth Au surfaces were prepared by depositing 50 nm of Au onto a clean Si wafer by thermal evaporation. No adhesion layer was used in this process. The 50 nm Au film was transferred to a cover glass by template stripping using NOA61 adhesive. The adhesive was cured under UV light for 5 minutes, with the TS Au film removed from the wafer using a razor blade directly prior to use.

**Patterned silicon wafer preparation.** Patterned silicon wafers were fabricated using FIB milling and plasma etching. A 50 nm-thick Au film was deposited onto a Si wafer by thermal evaporation. FIB milling (current = 0.79 nA) was employed to selectively remove the Au layer in a specific pattern. The removal of these regions exposed the underlying silicon, which was subsequently treated by plasma etching (100 W, $SF_6$ 25 sccm + $O_2$ 25 sccm, 5 minutes) to provide a roughened texture to the substrate. Finally, the residual Au film was removed using a gold etchant solution containing potassium iodide and iodine. The regions of the Si wafer initially milled by the FIB had a rough surface while those protected by Au had a flat surface, resulting in a patterned silicon wafer (Supplementary Note 13).

## Formation of 2D NAMs and 3D NAMs

**Synthesis of 2D AuNAMs and 3D AuNAMs.** 250 µL of AuNC stock solution was mixed with 250 µL DIW and 50 µL of 1 mM CTAB in a 1.5 mL microcentrifuge tube. After vortexing for 30 minutes, the solution was centrifuged (2,830 g, 10 minutes). The supernatant was then removed such that 25 µL of nanocube solution remained in the tube. For 2D and 3D assemblies, the nanocube solution was mixed with 20 µL DIW and 55 µL of 100 mM BDAC solution. The mixed solution was immediately dropcast onto a substrate. The spot volume varies depending on the specific application. Typically, 10 µL of solution was used for each spot. The substrates were then transferred to a humidity chamber and incubated at 25 °C. An optimized incubation time of 2 hours was used for representative 2D assembly on a flat Si wafer, TS Au film and a patterned Si wafer. An optimized incubation time of 3 hours was used for representative 3D assembly on borosilicate glass and silicon oxide film, and also for experiments in Fig. 4 performed on flat and rough Si wafers. The humidity chamber was prepared by laying damp laboratory wipes inside a petri dish, on which the substrate was placed. The petri dish lid was then closed, and the entire setup was sealed inside a zipper bag containing a small amount of water. After incubation the substrates were washed with 80 mM BDAC solution and left to dry under ambient conditions, with humidity less than 35%.

AuNC concentrations can be modified by changing the volume of AuNC stock solution; BDAC concentrations can be modified by changing the volume of the 100 mM BDAC solution and DIW when mixing the assembly solution while maintaining the total volume.

Experimental protocols for applying the MIDAS method to nanocubes of different sizes, Au octahedra, multicomponent AuNC/AgNC systems, and other depletants can be found in Supplementary Note 3.

**Synthesis of wafer-scale 2D AuNAMs**

The nanocube and depletant mixture described above was scaled up to a total volume of 21 mL, with final concentrations of 0.9 nM for the 50-nm AuNCs and 55 mM for BDAC. A 4-inch Si wafer placed inside a petri dish was fully immersed in the solution, the lid of the dish closed, and then incubated for 2 hours at 25 °C. The entire setup was sealed inside a zipper bag with a small amount of water. The wafer was washed with 80 mM BDAC solution after incubation and left to dry under ambient conditions.

**Characterization of 2D and 3D NAMs**

**Characterization of nanoparticles and micelles.** Characterization procedures for AuNCs using transmission electron microscopy, size measurements, calculation of nanocube concentration, and zeta potential measurements as well as characterization methods for depletion micelles are provided in Supplementary Note 2.

**SEM imaging.** Samples were dried over the course of a day, and washed carefully with DIW to remove excess surfactant residue on the surface before imaging. SEM images of 2D and 3D assembled structures were acquired with an Apreo 2 scanning electron microscope (Thermo Scientific).

**GI-SAXS measurement.** The GI-SAXS measurements of the 2D and 3D assembled structures were conducted with XEUSS 2.0 (Xenocs). The wavelength of the incident X-ray was 1.54189 Å (8.05 keV in photon energy), and the beam size was 0.7 mm x 0.4 mm. The sample-to-detector distance was 2500 mm and the data was collected for 1200 seconds. The lattice constant of assembled structures was calculated and used to calculate the size of lateral interparticle gaps by subtracting the edge length of the nanocubes.

**AFM measurement.** The AFM height maps were measured in tapping mode using the AFM unit in Ntegra (NT-MDT). The roughness of different substrates was measured. Also, measurement of the number of nanocube layers and the size of vertical gaps in 3D AuNAM was carried out by taking the edge length of the nanocubes into consideration. For measuring the patterned Si wafer, an NX-10 microscope (Park Systems) was used.

**UV-Vis-NIR reflectance spectroscopy measurement.** Reflectance spectra for 2D AuNAM and 3D AuNAM were measured with a Cary 5000 UV-Vis-NIR spectrophotometer (Agilent Technologies). The measurements were performed in the range of 400–2500 nm to observe the Mie-type resonant mode of periodic structures.

**FEM simulation.** Simulations were performed using a finite element method (FEM) in COMSOL Multiphysics software. Unit cells were designed considering the shape and size of nanocubes, interparticle gap size, and the crystal structure and number of layers in the assembled structure. Periodic boundary conditions in the x and y directions were applied to the unit cells. Air was designated as the surrounding medium, including in the nanogap region. Simulations resulted in reflectance spectra for the 2D and 3D AuNAMs. The reflectance spectrum for the 2D AuNAM was estimated from a monolayer configuration. For the 3D AuNAM, the reflectance spectrum was calculated by taking the spectra for different numbers of layers and weighting them according to their area percentage as estimated from AFM height maps. Weighted reflectance refers to the calculated spectrum of reflectance by being weighted by the surface coverage (Extended Data Fig. 5d) and summed up.

$$\text{Weighted reflectance (\%)} = \sum_{i=1} (\text{surface coverage of i}^{\text{th}} \text{ layer}) \times (\text{simulated reflectance of i}^{\text{th}} \text{ layer})$$

### *In situ* monitoring of assembly process

***In situ* optical microscopy setup.** The assembly process was monitored in real-time with an Olympus inverted microscope system (bright-field setup) using an oil immersion objective lens (UNPLAN, 100x, NA 1.4). The overall observation time was 3 hours to encompass the full process of nucleation, growth, termination and drying. In order to prevent solution evaporation, a silicone isolation chamber was affixed to the substrate, 60 µL of assembly solution added, and the chamber sealed with a cover glass. Due to the necessity of using a light-transmitting substrate, TS Au film was used for the observation of 2D assembly and borosilicate glass for 3D assembly. For further explanation of image processing and quantification of assembly rate, see Supplementary Note 7.

### SERS measurement

The sample was prepared by incubating a section of the wafer-scale 2D NAM in a 100 µM crystal violet solution. After 10 minutes, the wafer slices were removed and dried under $N_2$ gas. SERS measurements were performed with a Raman microscope (Renishaw inVia Qontor), using a 5× objective lens (N PLAN, NA=0.12) and a charge-coupled device (CCD) detector. A point map of the surface was obtained with a step size of 1 mm, using a 785 nm excitation laser at 4.29 mW power and 1 second acquisition time.

### Data Availability

The data supporting the findings of this study are available within this article and its supplementary information files.

Extended Figures and Supplementary Information include:

I. Extended Data Fig. 1 | Shape-based selective flocculation of AuNCs
II. Extended Data Fig. 2 | Effect of AuNC and BDAC concentrations on nucleation of AuNCs.
III. Extended Data Fig. 3 | AuNC assembly on various substrates.

**Author information**

**Authors and Affiliations**

Department of Chemistry, Seoul National University, Seoul 08826, South Korea

Yeonhee Lee, Seungsang Cha, Yuna Kwak & Jwa-Min Nam

Department of Chemistry, Stanford University, Stanford, California 94305, USA
Nicholas Juntunen & Grant M. Rotskoff

**Corresponding Authors**

Jwa-Min Nam - Department of Chemistry, Seoul National University, Seoul 08826, South Korea; https://orcid.org/0000-0002-7891-8482; Email: jmnam@snu.ac.kr

Grant M. Rotskoff - Department of Chemistry, Stanford University, Stanford, California 94305, USA; https://orcid.org/0000-0002-7772-5179; Email: rotskoff@stanford.edu

**Author Contributions**

Y. Lee, S. Cha, Y. Kwak and J.-M. Nam conceived the main idea and designed the experiments. Y. Lee, S. Cha and Y. Kwak performed the experiments and data analysis under the guidance of J.-M. Nam. S. Cha conducted the electromagnetic simulation and depletion interaction potential calculation under guidance of J.-M. Nam. N. Juntunen conducted the kinetic Monte Carlo simulations and mean field theory calculation under the guidance of G. M. Rotskoff. All authors contributed to the writing of the manuscript.

‡(Y. Lee, S. Cha, Y. Kwak, N. Juntunen) These authors contributed equally.

**Ethics declarations**

The authors declare no competing financial interests.

**Acknowledgements**

This research was supported by the National Research Foundation of Korea (NRF) grant funded by the Korea government (MSIT) (RS-2026-25469493) and the Nano·Material Technology Development Program through the National Research Foundation of Korea (NRF) funded by the Ministry of Science and ICT (RS-2024-00450828). The simulation results used resources of the National Energy Research Scientific Computing Center, a DOE Office of Science User Facility supported by the Office of Science of the U.S. Department of Energy under Contract No. DE-AC02-05CH11231 using NERSC award BES-ERCAP0033250. This material is based upon work supported by the National Science Foundation Graduate Research Fellowship under Grant No. DGE-2146755. G.M.R. was supported by the U.S. Department of Energy, Office of Science, Office of Basic Energy Sciences, under Award No. DE-SC0022917.

**Keywords**

Morphology-induced and depletion force-directed nanoparticle assembly on surface, nanoparticle self-assembly, nanoparticle arrays, nanoparticle supercrystals and superlattices, plasmonic metal nanoparticles